\documentclass[10pt,conference]{IEEEtran}
\usepackage{epsfig}
\usepackage{amsmath}
\usepackage{amssymb}
\usepackage{verbatim}
\usepackage{delarray}
\usepackage{graphics}
\usepackage{epic}

\newcommand{\inner}[2]{\langle{#1},{#2}\rangle}

\newcommand{\A}{{\mathcal{A}}}
\newcommand{\B}{{\mathcal{B}}}
\newcommand{\CC}{{\mathcal{C}}}

\newcommand{\FF}{{\mathcal{F}}}
\newcommand{\G}{{\mathcal{G}}}

\newcommand{\I}{{\mathcal{I}}}
\newcommand{\J}{{\mathcal{J}}}
\newcommand{\K}{{\mathcal{K}}}

\newcommand{\W}{{\mathcal{W}}}
\newcommand{\X}{{\mathcal{X}}}
\newcommand{\Y}{{\mathcal{Y}}}
\newcommand{\C}{{\mathbb{C}}}
\newcommand{\F}{{\mathbb{F}}}
\newcommand{\R}{{\mathbb{R}}}
\newcommand{\Z}{{\mathbb{Z}}}

\newcommand{\ab}{{\mathbf a}}

\newcommand{\Mb}{{\mathbf M}}

\newcommand{\ub}{{\mathbf u}}
\newcommand{\vb}{{\mathbf v}}
\newcommand{\wb}{{\mathbf w}}
\newcommand{\xb}{{\mathbf x}}
\newcommand{\yb}{{\mathbf y}}

\newcommand{\Ab}{{\mathbf A}}

\newcommand{\Xb}{{\mathbf X}}
\newcommand{\Yb}{{\mathbf Y}}

\newcommand{\ie}{{\em i.e., }}
\newcommand{\eg}{{\em e.g., }}

\newcommand{\etal}{\emph{et al.\ }}

\newcommand{\Tr}{\mathrm{Tr~}}

\newcommand{\openbox}{\leavevmode
     \hbox to.77778em{%
     \hfil\vrule
     \vbox to.675em{\hrule width.6em\vfil\hrule}%
     \vrule\hfil}}
\newcommand{\proofname}{Proof}

\newcommand{\qed}{\hspace*{1cm}\hspace*{\fill}\openbox}

\newcommand{\muleft}[1]{{\overleftarrow{\mu}_{\hskip-0.07cm #1}}}
\newcommand{\muright}[1]{{\overrightarrow{\mu}_{\hskip-0.07cm #1}}}

\newcommand{\Gleft}[1]{{\overleftarrow{\G}_{\hskip-0.07cm #1}}}
\newcommand{\Gright}[1]{{\overrightarrow{\G}_{\hskip-0.07cm #1}}}

\newcommand{\Kright}{{\overrightarrow{\K}}}

\newcommand{\Yleft}[1]{{\overleftarrow{Y}_{\hskip-0.10cm #1}}}
\newcommand{\Yright}[1]{{\overrightarrow{Y}_{\hskip-0.10cm #1}}}

\newcommand{\Ybright}{{\overrightarrow{\Yb}}}

\newcommand{\ybright}{{\overrightarrow{\yb}}}

\begin{document}

\title{Partition Functions of Normal Factor Graphs}

\author{\authorblockN{G. David Forney, Jr.}
\authorblockA{Laboratory for Information and Decision Systems\\
Massachusetts Institute of Technology\\
Cambridge, MA 02139, USA \\
forneyd@comcast.net}
\and
\authorblockN{Pascal O.~Vontobel}
\authorblockA{Hewlett--Packard Laboratories\\
1501 Page Mill Road\\
Palo Alto, CA 94304, USA\\
pascal.vontobel@ieee.org}}

\maketitle

\begin{abstract}

One of the most common types of functions in mathematics, physics, and
engineering is a sum of products, sometimes called a partition function.
After ``normalization,'' a sum of products has a natural graphical
representation, called a normal factor graph (NFG), in which vertices
represent factors, edges represent internal variables, and half-edges
represent the external variables of the partition function. In physics,
so-called trace diagrams share similar features.

We believe that the conceptual framework of representing sums of products as
partition functions of NFGs is an important and intuitive paradigm that,
surprisingly, does not seem to have been introduced explicitly in the
previous factor graph literature.

Of particular interest are NFG modifications that leave the partition function invariant.
A simple subclass of such NFG modifications offers a unifying view of the Fourier
transform, tree-based reparameterization, loop calculus, and  the Legendre transform.

\end{abstract}

\section{Introduction}
Functions that can be expressed as sums of products are ubiquitous in mathematics, science, and engineering.  Borrowing a physics term, we call such a function a \emph{partition function}.

In this paper, we will represent partition functions by \emph{normal factor graphs} (NFGs), which build on the concepts of factor graphs \cite{KFL01} and normal graphs \cite{F01}.  A factor graph represents a product of factors by a bipartite graph,  in which one set of vertices represents variables, while the other set of vertices represents factors.  By introducing ``normal'' degree restrictions as in \cite{F01}, we can represent a sum of products by an NFG in which edges represent variables and vertices represent factors.  Moreover, internal and external variables are distinguished in an NFG by being represented by edges of degree 2 and degree 1, respectively.
NFGs closely resemble the ``Forney-style
factor graphs'' (FFGs) of Loeliger \etal \cite{L04, L07}, with the difference
that ``closing the box'' (summing over internal variables) is always explicitly
assumed as part of the graph semantics.

There are as many applications of NFGs as there are of sums of products.  In this paper, we will present several applications that highlight the usefulness of the graphical approach:
\begin{itemize}
\item \textbf{Trace diagrams}, which are closely related to NFGs, often provide insight into linear algebraic relations, particularly of the kind that arise in various areas of physics;
\item The \textbf{sum-product algorithm} is naturally nicely derived in terms of NFGs;
\item The \textbf{normal factor graph duality theorem} \cite{AM10, F11} is a powerful general result, of which one corollary is the normal graph duality theorem of \cite{F01}.
\item The \textbf{holographic transformations} of NFGs of Al-Bashabsheh and Mao \cite{AM10}, which may be used to derive the ``holographic algorithms'' of Valiant \cite{V04} and others, may be further generalized to derive the ``tree-based reparameterization'' approach of Wainwright \etal \cite{WJW03}, the ``loop calculus'' results of Chertkov and Chernyak \cite{CC06, CC06b}, and the Lagrange duality results of Vontobel and Loeliger \cite{V02, VL03}.
\item \textbf{Linear codes defined on graphs} and their \textbf{weight generating functions} have natural representations as NFGs, as shown in \cite{F11}, but  we will not discuss this topic here.
\end{itemize} 

\section{Partition Functions and Graphs}

A \textbf{partition function} is any function $Z(\xb)$ that is given in ``sum-of-products form,'' as follows:
$$
Z(\xb) = \sum_{\yb \in \Y} \prod_{k \in \K} f_k(\xb_k, \yb_k), \quad \xb \in \X,
$$ 
where
\begin{itemize}
\item $\Xb$  is a set of $m$ \textbf{external variables} $X_i$ taking values $x_i$ in alphabets $\X_i, 1 \le i \le m$;
\item $\Yb$  is a set of $n$ \textbf{internal variables} $Y_j$ taking values $y_j$ in alphabets $\Y_i, 1 \le j \le n$;
\item each \textbf{factor} $f_k(\xb_k, \yb_k), k \in \K,$ is a function of certain subsets $\Xb_k \subseteq \Xb$ and $\Yb_k \subseteq \Yb$ of the sets of external and internal variables, respectively.
\end{itemize}
The set $\X = \prod_{i = 1}^m \X_i$ of all possible external variable configurations is called the \emph{domain} of the partition function, and the set $\Y = \prod_{j = 1}^n \Y_j$ of all possible internal variable configurations is called its \emph{configuration space}.
We say that a factor $f_k(\xb_k, \yb_k)$ \emph{involves} a variable $X_i$ (resp.\ $Y_j$) if $f_k$ is a function of that variable; \ie if $X_i \in \Xb_k$ (resp.\ $Y_j \in \Yb_k$).  For simplicity, we will assume that all functions are complex-valued, and that all variable alphabets are discrete.\footnote{Usually in physics a partition function is a sum over internal configurations (state configurations), and there are no external variables in our sense (although there may be parameters, such as temperature).  So our usage of ``partition function'' extends the usual terminology of physics.  Al-Bashabsheh and Mao \cite{AM10} use the term ``exterior function.''}

A particular sum-of-products form for a partition function will be called a \emph{realization}.  Different realizations that yield the same partition function $Z: \X \to \C$ will be called \emph{equivalent}.  We say that equivalent realizations \emph{preserve the partition function}.

\pagebreak

\subsection{Normal partition functions}

We will say that a realization of a partition function is \textbf{normal} if all external variables are involved in precisely one factor $f_k$, and all internal variables are involved in precisely two factors.  These degree restrictions were introduced in \cite{F01} in the context of behavioral graphs.

As observed in \cite{F01}, any realization may be converted to an equivalent normal realization by the following simple \textbf{normalization procedure}.  
\begin{itemize}
\item For every external variable $X_i$, if $X_i$ is involved in $p$ factors, then define $p$ \emph{replica variables} $X_{i\ell}, 1 \le \ell \le p$, replace $X_i$ by $X_{i\ell}$ in the $\ell$th factor in which $X_i$ is involved, and introduce one new factor, namely an \emph{equality indicator function} $\Phi_{=}(x_i, \{x_{i\ell}, 1 \le \ell \le p\})$ (see below).
\item For every internal variable $Y_j$, if $Y_j$ is involved in $q \ge 2$ factors, then define $q$ replica variables $Y_{j\ell}, 1 \le \ell \le q$, replace $Y_j$ by $Y_{j\ell}$ in the $\ell$th factor in which $Y_j$ is involved, and introduce one new factor, namely an equality indicator function $\Phi_{=}(\{y_{j\ell}, 1 \le \ell \le q\})$.  %Thus each replica variable $Y_{j\ell}$ becomes an internal variable that is involved in precisely two factors.  If $q=2$, then this conversion need not be performed.  If $q=1$, then multiply the partition function by a dummy factor $1(y_k)$ which is equal to 1 regardless of the value of $Y_k$.
\end{itemize}
Thus all replica variables are internal variables that are involved in precisely two factors, while the external variables $X_i$ become involved in only one factor, namely an equality indicator function.
Evidently this normalization procedure preserves the partition function. 

\subsection{Normal factor graphs}

For a normal realization of a partition function, a natural graphical model is a \textbf{normal factor graph} (NFG), in which vertices are associated with factors, ordinary edges (\ie hyperedges of degree 2) are associated with internal variables, ``half-edges'' \cite{F01} (\ie hyperedges of degree 1) are associated with external variables, and a variable edge or half-edge is incident on a factor vertex if the variable is involved in that factor.

\begin{figure}[h]
\setlength{\unitlength}{4pt}
\centering
\begin{picture}(70,5)(6, 3)
\multiput(16,5)(15,0){2}{\line(1,0){6}}
\put(37,3.5){\line(0,1){3}}
\put(18,6){$I$}
\put(33,6){$J$}
\put(7,2.5){\framebox(9,5){$w$}}
\put(22,2.5){\framebox(9,5){$M$}}
\put(43,5){$=$}
\put(52,2.5){\framebox(9,5){$v$}}
\put(61,5){\line(1,0){6}}
\put(67,3.5){\line(0,1){3}}
\put(63,6){$J$}
\end{picture}
\caption{Normal factor graph of a matrix multiplication $\vb = \wb M$.}
\label{Fig1}
\end{figure}
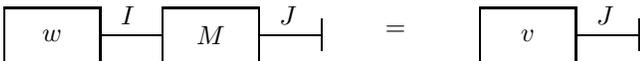

\vspace{1ex}
\noindent
\textbf{Example 1} (vector-matrix multiplication).
Consider a multiplication $\vb = \wb M$ of a vector $\wb$ by a matrix $M$, namely
$$v_j = \sum_{i \in \I} w_i M_{ij}, \quad j \in \J,$$ 
for some discrete index sets $\I$ and $\J$.  This may be interpreted as a normal realization of the function $v: \J \to \C$, with external variable $J$, internal variable $I$, and factors $w_i$ and $M_{ij}$.  Figure~\ref{Fig1} shows the corresponding normal factor graph, in which the vertices are represented by labeled boxes, and the half-edge is represented by a special dongle symbol.\footnote{The dongle symbol ``$\dashv$'' was chosen in \cite{F01} to suggest the possibility of a connection to another external half-edge in the manner of two railroad cars coupling, but of course this embellishment may be omitted.}  \qed

\subsection{Equality indicator functions}

We use special symbols for certain frequently occurring factors.  The most common and fundamental factor is the \textbf{equality indicator function} $\Phi_=$, which equals 1 if all incident variables (which must have a common alphabet) are equal, and equals 0 otherwise.  

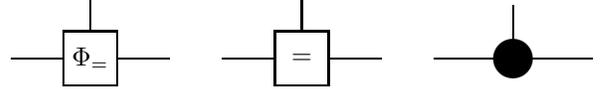
\begin{figure}[h]
\setlength{\unitlength}{4pt}
\centering
\begin{picture}(50,6)(3, 3)
\multiput(0,5)(20,0){2}{\line(1,0){5}}
\multiput(7.5,7.5)(20,0){2}{\line(0,1){3}}
\multiput(10,5)(20,0){2}{\line(1,0){5}}
\put(5,2.5){\framebox(5,5){$\Phi_=$}}
\put(25,2.5){\framebox(5,5){$=$}}
\put(47.5,5){\circle*{5}}
\put(40,5){\line(1,0){6}}
\put(47.5,5){\line(0,1){5}}
\put(49,5){\line(1,0){6}}
\end{picture}
\caption{Three representations of an equality indicator function of degree 3.}
\label{Fig2}
\end{figure}

Figure~\ref{Fig2} shows three ways of representing an equality indicator function:  first, by a vertex labeled by $\Phi_=$; second, by a vertex labeled simply by an equality sign $=$;  and third, as a junction vertex.  The second representation makes a connection with the behavioral graph literature (\eg Tanner graphs), where vertices represent constraints rather than factors.  The third representation makes connections with ordinary block diagrams, where any number of edges representing the same variable may meet at a junction, as well as with the factor graph literature, where variables are represented by vertices rather than by edges.

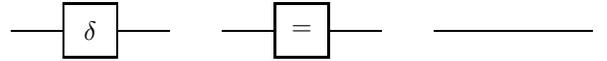
\begin{figure}[h]
\setlength{\unitlength}{4pt}
\centering
\begin{picture}(50,4)(3, 3)
\multiput(0,5)(20,0){2}{\line(1,0){5}}
\multiput(10,5)(20,0){2}{\line(1,0){5}}
\put(5,2.5){\framebox(5,5){$\delta$}}
\put(25,2.5){\framebox(5,5){$=$}}
\put(40,5){\line(1,0){15}}
\end{picture}
\caption{Three representations of an equality indicator function of degree 2.}
\label{Fig2a}
\end{figure}

An equality indicator function of degree 2 is often denoted by a Kronecker delta function $\delta$.  Since such a function connects only two edges and constrains their respective variables to be equal, it may simply be omitted, as shown in Figure~\ref{Fig2a}.\footnote{The last equivalence shown in Figure~\ref{Fig2a} is actually a bit problematic, since a single edge  is not a legitimate normal factor graph;  however, as a component of a normal factor graph, such an edge is always incident on some factor vertex $f_k$, and since the combination of a factor $f_k$ involving some internal variable $Y_j$ with an equality function $\Phi_=(y_j, y_j')$ is just the same factor with $Y_j'$ substituted for $Y_j$, this substitution can be made in any legitimate NFG (see also \cite{AM10}).}

\section{Trace Diagrams}

It turns out that physicists have long used graphical diagrams called ``trace diagrams'' \cite{C08, MP09, P05, P09, S90} that use semantics similar to those of NFGs.  In this section we give a brief exposition of this topic, following \cite{P09}.  %(We are grateful to Pascal Vontobel for first noticing this connection.)

\begin{figure}[h]
\setlength{\unitlength}{4pt}
\centering
\begin{picture}(40,7)(-2, 2)
\multiput(0,5)(20,0){2}{\line(1,0){5}}
\multiput(10,5)(20,0){2}{\line(1,0){5}}
\put(5,2.5){\framebox(5,5){$M$}}
\put(25,2.5){\framebox(5,5){$M$}}
\put(6,0){(a)}
\put(2,6){$I$}
\put(12,6){$J$}
\put(26,0){(b)}
\put(22,6){$I$}
\put(20,5){\line(0,1){5}}
\put(20,10){\line(1,0){15}}
\put(35,5){\line(0,1){5}}
\end{picture}
\caption{Representations of (a) a matrix $M$;  (b) the trace of $M$.}
\label{Fig3}
\end{figure}
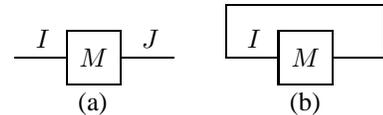

In trace diagrams, the factors are often vectors, matrices, tensors, and so forth, and the variables are typically their indices.  For instance, a matrix $M= \{M_{ij}, i \in \I, j \in \J\}$ may be considered to be a function of the two variables $I$ and $J$, and is represented as a vertex with two incident edges, as in Figure~\ref{Fig3}(a).

Trace diagrams use the NFG convention that dangling edges (half-edges) represent external variables, whereas ordinary edges
represent internal variables, and are to be summed over.  For example, if the
matrix $M$ is square (\ie the index alphabets $\I$ and $\J$ are the same), and
the half-edges representing $I$ and $J$ are connected as in
Figure~\ref{Fig3}(b), then the resulting figure represents the \emph{trace} of
$M$, since $\Tr M = \sum_{i} M_{ii}$.  This apparently explains why these
kinds of graphical models are known as ``trace diagrams.''

The convention that indices that appear twice are implicitly to be summed over is known in physics as the \emph{Einstein summation convention}. This convention is used rather generally in physics, not just with trace diagrams.

Trace diagrams permit visual proofs of various relationships in linear algebra.  For example, Figure~\ref{Fig3a} proves the identity $\Tr ABC = \Tr BC\!A$.

\begin{figure}[h]
\setlength{\unitlength}{4pt}
\centering
\begin{picture}(40,16)(0, 4)
\multiput(0,15)(10,0){4}{\line(1,0){5}}
\put(5,12.5){\framebox(5,5){$A$}}
\put(15,12.5){\framebox(5,5){$B$}}
\put(25,12.5){\framebox(5,5){$C$}}
\put(35,15){\line(0,1){5}}
\put(0,20){\line(1,0){35}}
\put(0,15){\line(0,1){5}}
\put(-5,5){$=$}
\multiput(0,5)(10,0){4}{\line(1,0){5}}
\put(5,2.5){\framebox(5,5){$B$}}
\put(15,2.5){\framebox(5,5){$C$}}
\put(25,2.5){\framebox(5,5){$A$}}
\put(35,5){\line(0,1){5}}
\put(0,10){\line(1,0){35}}
\put(0,5){\line(0,1){5}}
\end{picture}
\caption{Proof of the identity $\Tr ABC = \Tr BC\!A$.}
\label{Fig3a}
\end{figure}

If $\ub$ and $\vb$ are two real vectors with a common index set $\I$, then their \emph{dot product} (inner product) is defined as 
$$
\ub \cdot \vb = \sum_{i\in \I} u_i v_i.
$$ 
The trace diagram (or normal factor graph) of a dot product is illustrated in Figure~\ref{Fig4}(a).

\begin{figure}[h]
\setlength{\unitlength}{4pt}
\centering
\begin{picture}(40,7)(3, 1)
\put(5,5){\line(1,0){5}}
\put(7,6){$I$}
\put(0,2.5){\framebox(5,5){$\ub$}}
\put(10,2.5){\framebox(5,5){$\vb$}}
\put(20,2.5){\framebox(5,5){$\ub$}}
\put(25,5){\line(1,0){5}}
\put(30,2.5){\framebox(5,5){$\varepsilon$}}
\put(35,5){\line(1,0){5}}
\put(40,2.5){\framebox(5,5){$\vb$}}
\put(32.5,7.5){\line(0,1){3.5}}
\put(31,11){\line(1,0){3}}
\put(33,8.5){$I$}
\put(27,6){$J$}
\put(37,6){$K$}
\put(6,0){(a)}
\put(31,0){(b)}
\end{picture}
\caption{Representations of (a) a dot product $\ub \cdot \vb$;  (b) a cross product $\ub \times \vb$.}
\label{Fig4}
\end{figure}

If $\ub$ and $\vb$ are two real three-dimensional vectors, then their \emph{cross product} $\ub \times \vb = \wb$ is defined by 
\begin{align*}
w_1 &= u_2 v_3 - u_3 v_2; \\
w_2 &= u_3 v_1 - u_1 v_3; \\
w_3 &= u_1 v_2 - u_2 v_1.
\end{align*}
Equivalently,
$$
w_i = \sum_{j=1}^3 \sum_{k=1}^3 \varepsilon_{ijk} u_j v_k,
$$
where we use the \emph{Levi-Civita symbol} $\varepsilon_{ijk}$, defined as
$$
\varepsilon_{ijk} 
  = \begin{cases}
      +1, & \text{if $ijk$ is an even permutation of $123$;} \\
      -1, & \text{if $ijk$ is an odd permutation of 123;} \\
      \ \ 0,  & \text{otherwise}. 
    \end{cases}
$$

\pagebreak
Thus $\wb$ is given in the form of a normal partition function with external variable $I$ and internal variables $J$ and $K$.
The trace diagram or NFG of this cross product is illustrated in Figure~\ref{Fig4}(b).
(Notice that in this case the order of the indices is important, since $\varepsilon_{ijk} = - \varepsilon_{jik}$.)

Similarly, the determinant of a $3\times 3$ matrix $M$ may be written in terms of $\varepsilon_{ijk}$ as 
$$
\det M =  \sum_{i=1}^3 \sum_{j=1}^3 \sum_{k=1}^3 \varepsilon_{ijk} M_{1i} M_{2j} M_{3k}.
$$
Thus if $\Mb_1, \Mb_2$ and $\Mb_3$ are the three rows of $M$, then its determinant
may be represented in trace diagram or normal factor graph notation as in Figure~\ref{Fig5}.

\begin{figure}[h]
\setlength{\unitlength}{4pt}
\centering
\begin{picture}(40,15)(15, 3)
\put(20,2.5){\framebox(5,5){$\Mb_2$}}
\put(25,5){\line(1,0){5}}
\put(30,2.5){\framebox(5,5){$\varepsilon$}}
\put(35,5){\line(1,0){5}}
\put(40,2.5){\framebox(5,5){$\Mb_3$}}
\put(32.5,7.5){\line(0,1){5}}
\put(30,12.5){\framebox(5,5){$\Mb_1$}}
\put(33,9.5){$I$}
\put(27,6){$J$}
\put(37,6){$K$}
\end{picture}
\caption{Representation of a determinant $\det \{\Mb_1, \Mb_2, \Mb_3\}$. }
\label{Fig5}
\end{figure}

Figure~\ref{Fig5} shows that the determinant of $M$ may be expressed in three equivalent ways, as follows:
\begin{eqnarray*}
\det M & = & 
\Mb_1 \cdot (\Mb_2 \times \Mb_3) \\ & = & \Mb_2 \cdot (\Mb_3 \times \Mb_1) \\ & = & \Mb_3 \cdot (\Mb_1 \times \Mb_2).
\end{eqnarray*}

The trace diagram notation permits other operations that have not heretofore been considered in the factor graph literature.  For example, two trace diagrams with the same sets of external variables that are connected by a plus or minus sign represent the \emph{sum} or \emph{difference} of the corresponding partition functions.\footnote{A product of partition functions is represented simply by a disconnected factor graph, with each component graph representing a component function.}
For example, Figure~\ref{Fig6} illustrates the ``contracted epsilon identity,'' namely
$$
\sum_{k=1}^3 \varepsilon_{ijk}\varepsilon_{k\ell m} = \delta_{i\ell}\delta_{jm} - \delta_{im}\delta_{j\ell}.
$$

\begin{figure}[h]
\setlength{\unitlength}{4pt}
\centering
\begin{picture}(65,5)(0, 1)
\put(2,1){\line(1,0){3}}
\put(2,4){\line(1,0){3}}
\put(2,0){\line(0,1){2}}
\put(2,3){\line(0,1){2}}
\put(3,4.5){$I$}
\put(3,1.5){$J$}
\put(5,0){\framebox(5,5){$\varepsilon$}}
\put(10,2.5){\line(1,0){5}}
\put(12,3){$K$}
\put(15,0){\framebox(5,5){$\varepsilon$}}
\put(20,1){\line(1,0){3}}
\put(20,4){\line(1,0){3}}
\put(23,0){\line(0,1){2}}
\put(23,3){\line(0,1){2}}
\put(20.5,4.5){$M$}
\put(21.5,1.5){$L$}
\put(25.5,2.5){$=$}
\put(30,1){\line(4,1){12}}
\put(30,4){\line(4,-1){12}}
\put(30,0){\line(0,1){2}}
\put(30,3){\line(0,1){2}}
\put(42,0){\line(0,1){2}}
\put(42,3){\line(0,1){2}}
\put(31,4.5){$I$}
\put(31,1.5){$J$}
\put(39,4.5){$M$}
\put(40,1.5){$L$}
\put(45,2.5){$-$}
\put(50,1){\line(1,0){12}}
\put(50,4){\line(1,0){12}}
\put(50,0){\line(0,1){2}}
\put(50,3){\line(0,1){2}}
\put(62,0){\line(0,1){2}}
\put(62,3){\line(0,1){2}}
\put(51,4.5){$I$}
\put(51,1.5){$J$}
\put(59,4.5){$M$}
\put(60,1.5){$L$}
\end{picture}
\caption{Contracted epsilon identity. }
\label{Fig6}
\end{figure}

From this identity, or its corresponding trace diagram, we can derive such
identities as
\begin{align*}
  (\ub \times \vb) \times \wb
    &= (\ub \cdot \wb) \vb -  (\vb \cdot \wb)\ub,
\end{align*}
illustrated in Figure~\ref{Fig7}(a), or
\begin{align*}
  (\ub \times \vb) \cdot (\wb \times \xb)
    &= (\ub \cdot \wb) (\vb \cdot \xb) - (\ub \cdot \xb) (\vb \cdot \wb),
\end{align*}
illustrated in Figure~\ref{Fig7}(b), which reduce expressions involving two
cross products to simpler forms involving only dot products.

\vspace{3ex}

\begin{figure}[h]
\setlength{\unitlength}{4pt}
\centering
\begin{picture}(65,15)(2, -1)
\put(25,7.5){(a)}
\put(0,10){\framebox(2,2){$\vb$}}
\put(0,13){\framebox(2,2){$\ub$}}
\put(2,11){\line(1,0){3}}
\put(2,14){\line(1,0){3}}
\put(5,10){\framebox(5,5){$\varepsilon$}}
\put(10,12.5){\line(1,0){5}}
\put(15,10){\framebox(5,5){$\varepsilon$}}
\put(20,11){\line(1,0){3}}
\put(20,14){\line(1,0){3}}
\put(23,13){\line(0,1){2}}
\put(23,10){\framebox(2,2){$\wb$}}
\put(25.5,12.5){$=$}
\put(28,10){\framebox(2,2){$\vb$}}
\put(28,13){\framebox(2,2){$\ub$}}
\put(30,11){\line(4,1){12}}
\put(30,14){\line(4,-1){12}}
\put(42,13){\line(0,1){2}}
\put(42,10){\framebox(2,2){$\wb$}}
\put(45,12.5){$-$}
\put(50,11){\line(1,0){12}}
\put(50,14){\line(1,0){12}}
\put(48,10){\framebox(2,2){$\vb$}}
\put(48,13){\framebox(2,2){$\ub$}}
\put(62,10){\framebox(2,2){$\wb$}}
\put(62,13){\line(0,1){2}}
\put(25,-2.5){(b)}
\put(0,0){\framebox(2,2){$\vb$}}
\put(0,3){\framebox(2,2){$\ub$}}
\put(2,1){\line(1,0){3}}
\put(2,4){\line(1,0){3}}
\put(5,0){\framebox(5,5){$\varepsilon$}}
\put(10,2.5){\line(1,0){5}}
\put(15,0){\framebox(5,5){$\varepsilon$}}
\put(20,1){\line(1,0){3}}
\put(20,4){\line(1,0){3}}
\put(23,0){\framebox(2,2){$\wb$}}
\put(23,3){\framebox(2,2){$\xb$}}
\put(25.5,2.5){$=$}
\put(28,0){\framebox(2,2){$\vb$}}
\put(28,3){\framebox(2,2){$\ub$}}
\put(30,1){\line(4,1){12}}
\put(30,4){\line(4,-1){12}}
\put(42,0){\framebox(2,2){$\wb$}}
\put(42,3){\framebox(2,2){$\xb$}}
\put(45,2.5){$-$}
\put(50,1){\line(1,0){12}}
\put(50,4){\line(1,0){12}}
\put(48,0){\framebox(2,2){$\vb$}}
\put(48,3){\framebox(2,2){$\ub$}}
\put(62,0){\framebox(2,2){$\wb$}}
\put(62,3){\framebox(2,2){$\xb$}}
\end{picture}
\caption{Cross product identities:  (a) $(\ub \times \vb) \times \wb = (\ub \cdot \wb) \vb -  (\vb \cdot \wb)\ub$;  (b) $(\ub \times \vb) \cdot (\wb \times \xb) = (\ub \cdot \wb) (\vb \cdot \xb) - (\ub \cdot \xb) (\vb \cdot \wb)$. }
\label{Fig7}
\end{figure}

\section{The sum-product algorithm}

The sum-product algorithm is an efficient method for computing partition functions of cycle-free graphs. It has been explained many times, including in \cite{F01}.  Here we explain it again in the language of normal factor graphs, with the objective of achieving a clearer and more intuitive explanation than in \cite{F01}.  We freely use ideas from \eg $\cite{AM00, KFL01, L04, L07, WLK95}$.
  
As Al-Bashabsheh and Mao \cite{AM10} have emphasized, a partition function is
completely determined by the set $\{f_k(\xb_k, \yb_k)\}$ of factors,
independent of their ordering.  In evaluating a partition function, factors
may be arbitrarily ordered and grouped.  This observation (called the
``generalized distributive law'' by Aji and McEliece \cite{AM00}) is at the
root of the sum-product algorithm.

We start with a normal realization of a partition function with \emph{no
  external variables} whose associated normal graph $\G$ is \emph{connected}
and \emph{cycle-free}.  Thus the partition function of $\G$ is a constant,
denoted by $Z(\G)$, and $\G$ is an ordinary graph (no half-edges) that
moreover is a tree.

\begin{figure}[h]
\setlength{\unitlength}{4pt}
\centering
\begin{picture}(50,4)(0, 1)
\put(-5,0){\framebox(5,5){$\G$}}
\put(1.5,2){$=$}
\put(5,0){\framebox(5,5){$\Gright{j}$}}
\put(10,2.5){\line(1,0){10}}
\put(14,3.5){$Y_j$}
\put(20,0){\framebox(5,5){$\Gleft{j}$}}
\put(29,2.5){$\Rightarrow$}
\put(35,0){\framebox(5,5){$\Gright{j}$}}
\put(40,2.5){\line(1,0){4}}
\put(44,1){\line(0,1){3}}
\put(40.5,3.5){$\Yright{j}$}
\put(46,2.5){\line(1,0){4}}
\put(46,1){\line(0,1){3}}
\put(46.5,3.5){$\Yleft{j}$}
\put(50,0){\framebox(5,5){$\Gleft{j}$}}
\end{picture}
\caption{Disconnecting a cycle-free NFG $\G$ by a cut through edge $Y_j$. }
\label{Fig9}
\end{figure}
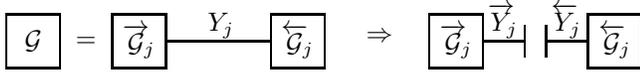

A connected graph $\G$ is cycle-free if and only if any \emph{cut} through any edge  $Y_j$ divides $\G$ into two disconnected graphs,
which we label arbitrarily as $\Gright{j}$ and $\Gleft{j}$. Such a cut divides
the edge associated with $Y_j$ into two half-edges associated with two
external variables, denoted by $\Yright{j}$ and $\Yleft{j}$, with the same
alphabet $\Y_j$ as $Y_j$, as illustrated in Figure~\ref{Fig9}.

Let us define the \emph{messages} $\muright{j}(y_{j})$ and $\muleft{j}(y_{j})$
as the partition functions of $\Gright{j}$ and $\Gleft{j}$, respectively; \ie
$$
\muright{j}(y_{j}) = \sum_{\ybright \in \Ybright}
\prod_{k \in \Kright} f_k(\yb_k),
$$
where $\Ybright$ is the set of left-side variables (excluding $Y_j$), and
$\Kright$ is the set of indices of left-side factors, and similarly for
$\muleft{j}(y_{j})$.  The goal of the sum-product algorithm is to compute the
messages $\muright{j}(y_{j}), \muleft{j}(y_{j})$ for every internal
variable $Y_j$.

\begin{figure}[h]
\setlength{\unitlength}{4pt}
\centering
\begin{picture}(50,13)(-22, 1)
\put(-25,5){\framebox(5,5){$\Gright{j}$}}
\put(-20,7.5){\line(1,0){5}}
\put(-15,6){\line(0,1){3}}
\put(-19,8.5){$\Yright{j}$}
\put(-10,7.5){$=$}
\put(-5,0){\framebox(8,5){$\Gright{j''}$}}
\put(3,2.5){\line(1,0){7}}
\put(5,3.5){$Y_{j''}$}
\put(-5,10){\framebox(8,5){$\Gright{j'}$}}
\put(3,12.5){\line(1,0){7}}
\put(5,13.5){$Y_{j'}$}
\put(10,0){\framebox(15,15){$f(y_j, y_{j'}, y_{j''})$}}
\put(25,7.5){\line(1,0){5}}
\put(30,6){\line(0,1){3}}
\put(26,8.5){$\Yright{j}$}
\end{picture}
\caption{Expressing an NFG  in terms of subgraphs connected to a vertex.}
\label{Fig12}
\end{figure}
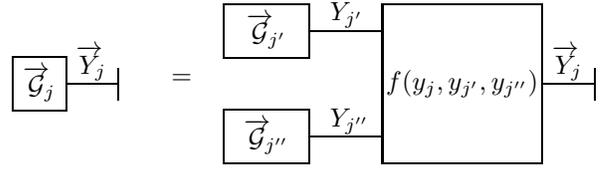

To compute a message such as $\muright{j}(y_j)$, consider the factor vertex to
which $\Yright{j}$ is attached.  For simplicity, let us suppose that this
vertex has degree 3, and that the associated factor is $f(y_j, y_{j'},
y_{j''})$, as shown in Figure~\ref{Fig12}.

Since $\G$ is cycle-free, the subgraphs $\Gright{j'}$ and $\Gright{j''}$ that
extend from the edges $Y_{j'}$ and $Y_{j''}$ must be disjoint.  Their
partition functions, $\muright{j'}(y_{j'})$ and $\muright{j''}(y_{j''})$,
include all factors in $\muright{j}(y_j)$ except $f(y_j, y_{j'}, y_{j''})$,
and sum over all internal variables except $Y_{j'}$ and $Y_{j''}$. Therefore
the partition function $\muright{j}(y_j)$ of $\Gright{j}$ may be
expressed in terms of the partition functions of these subgraphs as follows:
$$
\muright{j}(y_j) 
 %& = & \sum_{y_{j'} \in \Y_{j'}} \sum_{y_{j''} \in \Y_{j''}}  f(y_j, y_{j'}, y_{j''}) \left(\sum_{\yb' \in \Yb'}  \prod_{k \in \K'} f_k(\yb_k) \right) \left( \sum_{\yb'' \in\Yb''}\prod_{k \in \K''} f_k(\yb_k) \right) \\
= \sum_{y_{j'} \in \Y_{j'}} \sum_{y_{j''} \in \Y_{j''}}  f(y_j, y_{j'}, y_{j''}) \muright{j'}(y_{j'}) \muright{j''}(y_{j''}).
$$
More generally, if the factor vertex to which edge $Y_j$ is attached is
$f_k(\yb_k)$, then the message update rule is
$$
\muright{j}(y_j) = \sum_{\yb_k \setminus\{y_j\}} f_k(\yb_k) \prod_{j' \in \J_{k}\setminus \{j\}} \muright{j'}(y_{j'}).
$$
This is called the \textbf{sum-product update rule}.
 
Since $\G$ is connected and cycle-free, it is a tree (assuming that it is finite).  Each
message $\muright{j}$ has a \emph{depth} equal to the maximum
length of any path from that message to any leaf vertex.  The messages at depth
$1$ can be computed immediately, the messages at depth 2 can be computed as
soon as the messages at depth $1$ are known, and so forth.  If $\G$ is finite,
then all messages can be computed in at most $\delta(\G)$ rounds, where
$\delta(\G)$ is the maximum possible depth, called the \emph{diameter}.

For any internal variable $Y_j$, we define the \textbf{marginal partition
  function} $Z_j(y_j)$ as
$$Z_j(y_j) =\muright{j}(y_j) \muleft{j}(y_j), \quad y_j \in \Y_j.$$
Thus $Z_j(y_j)$ is simply the componentwise (dot) product of the messages
$\muright{j}(y_j)$ and $\muleft{j}(y_j)$.  This is
sometimes called the \textbf{past-future decomposition rule} \cite{F01}.

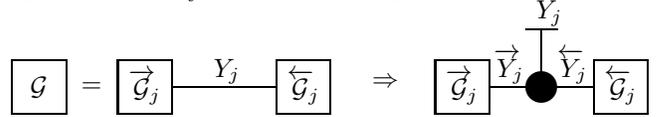
\begin{figure}[h]
\setlength{\unitlength}{4pt}
\centering
\begin{picture}(50,7)(0, 1)
\put(-5,0){\framebox(5,5){$\G$}}
\put(1.5,2){$=$}
\put(5,0){\framebox(5,5){$\Gright{j}$}}
\put(10,2.5){\line(1,0){10}}
\put(14,3.5){$Y_j$}
\put(20,0){\framebox(5,5){$\Gleft{j}$}}
\put(29,2.5){$\Rightarrow$}
\put(35,0){\framebox(5,5){$\Gright{j}$}}
\put(40,2.5){\line(1,0){4}}
\put(45,4){\line(0,1){4}}
\put(43.5,8){\line(1,0){3}}
\put(44.5,9){$Y_{j}$}
\put(45,2.5){\circle*{3}}
\put(40.5,3.5){$\Yright{j}$}
\put(46,2.5){\line(1,0){4}}
\put(46.5,3.5){$\Yleft{j}$}
\put(50,0){\framebox(5,5){$\Gleft{j}$}}
\end{picture}
\caption{Converting $Y_j$ from internal to external by inserting a ``tap.''}
\label{Fig10}
\end{figure}

Graphically, $Z_j(y_j)$ is the partition function of the graph obtained from
$\G$ by converting $Y_j$ from an internal to an external variable as shown in
Figure~\ref{Fig10}; \ie by replacing the edge associated with $Y_j$ by a
``tap'' consisting of the concatenation of an edge labeled by
$\Yright{j}$, an equality indicator function, and another edge
labeled by $\Yleft{j}$, with a further half-edge labeled by $Y_j$
attached to the equality indicator function.

Conversely, $Z(\G)$ is the partition function of the graph obtained by
converting $Y_j$ back to an internal variable; \ie by summing $Z_j(y_j)$ over
$Y_j$:
$$
Z(\G) = \sum_{y_j \in \Y_j} Z_j(y_j) = \sum_{y_j \in \Y_j} \muright{j}(y_j) \muleft{j}(y_j).
$$
Thus, for any edge $Y_j$, $Z(\G)$ is simply the dot product of the messages
$\muright{j}$ and $\muleft{j}$.

\section{Holographic Transformations}

In this section, we recapitulate and generalize the concept of ``holographic
transformations'' of normal factor graphs, which was introduced by
Al-Bashabsheh and Mao \cite{AM10}, and their ``generalized Holant theorem,''
which relates the partition function of a normal factor graph to that of its
holographic transform.  This theorem generalizes the Holant theorem of Valiant
\cite{V04} (see also \cite{CC07, CL07, CLX08, CLX10, V10}), which has been used to
show that some seemingly intractable counting problems on graphs are in fact
tractable.

Using this concept, Al-Bashabsheh and Mao \cite{AM10} were able to prove a very general and powerful Fourier transform duality theorem for normal factor graphs, of which the original normal graph duality theorem of \cite{F01} is an immediate corollary.  We give a variation of this proof which is perhaps even simpler (compare also the proof in \cite{F11}).

In the last section of this paper, we will sketch further applications of this general approach. 

\subsection{General approach}

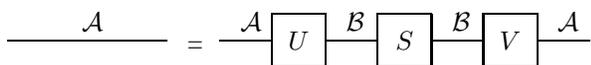
\begin{figure}[h]
\setlength{\unitlength}{4pt}
\centering
\begin{picture}(50,4)(3, 3)
\put(0,5){\line(1,0){15}}
\put(7,6){$\A$}
\put(17,4){=}
\put(20,5){\line(1,0){5}}
\put(22,6){$\A$}
\put(25,2.5){\framebox(5,5){$U$}}
\put(30,5){\line(1,0){5}}
\put(32,6){$\B$}
\put(35,2.5){\framebox(5,5){$S$}}
\put(40,5){\line(1,0){5}}
\put(42,6){$\B$}
\put(45,2.5){\framebox(5,5){$V$}}
\put(50,5){\line(1,0){5}}
\put(52,6){$\A$}
\end{picture}
\caption{A concatenation of factors that is equivalent to the identity.}
\label{FigUSV}
\end{figure}

The general approach can be explained very simply, as follows.  Let $\A$ and
$\B$ be two finite alphabets, which will often be of the same size; \ie $|\A|
= |\B|$.  Let $U(a,b), S(b,b')$, and $V(b',a')$ be complex-valued factors
involving variables $A$, $B$, $B'$, and $A'$ defined on $\A$, $\B$, $\B$, and
$\A$, respectively; alternatively, we may regard $U, S$, and $V$ as matrices.
Finally, suppose that the concatenation $U\!SV$, shown in Figure~\ref{FigUSV},
is the identity factor $\delta_{aa'}$, which can be represented simply as an
ordinary edge as in Figure~\ref{Fig2a}.\footnote{Here and subsequently we may label an internal edge simply by its alphabet, without introducing dummy internal variables.}

We then have the following obvious lemma:

\vspace{1ex}
\noindent
\textbf{Lemma} (generalized holographic transformations). In any NFG, any ordinary edge may be replaced by a concatenation of factors  $U\!SV$ equivalent to the identity, as in Figure~\ref{FigUSV}, without changing the partition function.  \qed \vspace{1ex}

The ``holographic transformations'' of \cite{AM10} involve similar replacements, except without the middle factor $S$ (alternatively, with $S(b,b') = \delta_{bb'}$).  Al-Bashabsheh and Mao \cite{AM10} call $\B$ the \emph{coupling alphabet}, and say that $U$ and $V$ are \emph{dual} with respect to $\B$.  When $|\A| = |\B|$, they say that $U$ and $V$ are \emph{transformers};  in this case, as matrices, $U$ and $V$ are inverses.

If a normal factor graph has external variables $X_i$, then they may be transformed as well, by the insertion of a factor or matrix $W_i(x_i,w_i)$ defined on $\X_i \times \W_i$, where $\W_i$ is the alphabet of a transformed external variable $W_i$.  Thus the partition function is transformed into a function of the new external variables $W_i$.  This is the essence of the ``generalized Holant theorem'' of \cite{AM10}.  (The original Holant theorem of Valiant \cite{V04} applies when there are no external variables.)

\subsection{General normal factor graph duality theorem}

This general approach yields a very simple proof of the ``general normal factor graph duality theorem'' of \cite{AM10, F11}.

Suppose that we have a normal factor graph in which each variable alphabet $\A$ is a finite-dimensional vector space over a finite field $\F$ of characteristic $p$ (\ie $p$ is the least positive integer such that $p\alpha = 0$ for all $\alpha \in \F$).  The dual space $\hat{\A}$ is then a vector space over $\F$ of the same dimension as $\A$, and there is a well-defined $\Z_p$-valued \emph{inner product} $\inner{\hat{a}}{a}$ with the usual properties;  \eg $\inner{\hat{a}}{0} = \inner{0}{a} = 0$, $\inner{\hat{a}}{a + a'} = \inner{\hat{a}}{a} + \inner{\hat{a}}{a'}$, and so forth (see, \eg  \cite{F98}).  

Given a complex-valued function $f: \A \to \C$ defined on $\A$, its \emph{Fourier transform} is then defined as the complex-valued function $F: \hat{\A} \to \C$ on $\hat{A}$ that maps $\hat{a}$ to
$$
F(\hat{a}) = \sum_{a \in \A} f(a)  \omega^{\inner{\hat{a}}{a}}, \quad \hat{a} \in \hat{\A},
$$
where $\omega = e^{2\pi i/p}$ is a primitive complex $p$th root of unity.

\begin{figure}[h]
\setlength{\unitlength}{4pt}
\centering
\begin{picture}(50,5)(11, 3)
\multiput(16,5)(15,0){2}{\line(1,0){6}}
\put(37,3.5){\line(0,1){3}}
\put(18,6){$\A$}
\put(33,6){$\hat{\A}$}
\put(7,2.5){\framebox(9,5){$f$}}
\put(22,2.5){\framebox(9,5){$\FF_{\A}$}}
\put(43,5){$=$}
\put(52,2.5){\framebox(9,5){$F$}}
\put(61,5){\line(1,0){6}}
\put(67,3.5){\line(0,1){3}}
\put(63,6){$\hat{\A}$}
\end{picture}
\caption{Normal factor graph of a Fourier transform.}
\label{Fig17}
\end{figure}
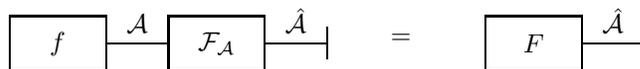

In an NFG, a Fourier transform may be represented as in Figure~\ref{Fig17}, where the Fourier transform factor is
$$\FF_{\A} = \{\omega^{\inner{\hat{a}}{a}}: \hat{a} \in \hat{\A}, a \in \A\}.$$
The transform $F(\hat{a})$ is obtained by summing over $\A$, which in this case amounts to a matrix-vector multiplication.  

Note that as a factor in an NFG, we do not have to distinguish between $\FF_{\A}$ and its transpose;  $\FF_{\A}$ is simply a function of the two variables corresponding to the two incident edges, and as a matrix can act on either variable.  Thus $\FF_{\A}$ can act also as a Fourier transform $\FF_{\hat{\A}}$ on a function of $\hat{\A}$.

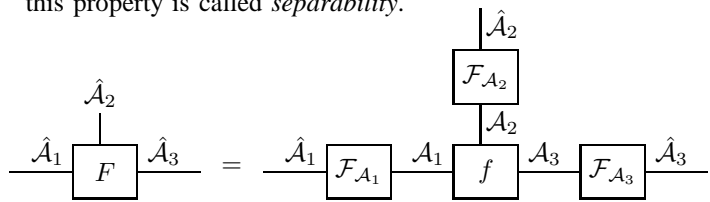
\begin{figure}[h]
\setlength{\unitlength}{4pt}
\centering
\begin{picture}(50,14)(9, 3)
\multiput(1,5)(12,0){2}{\line(1,0){6}}
\multiput(9.5,7.5)(12,0){1}{\line(0,1){3}}
\put(3,6){$\hat{\A}_1$}
\put(14,6){$\hat{\A}_{3}$}
\put(8,11.5){$\hat{\A}_2$}
\put(7,2.5){\framebox(6,5){$F$}}
\put(21,5){$=$}
\multiput(25,5)(12,0){4}{\line(1,0){6}}
\multiput(45.5,7.5)(0,8){1}{\line(0,1){4}}
\multiput(45.5,16.5)(0,8){1}{\line(0,1){4}}
\put(27,6){$\hat{\A}_1$}
\put(39,6){$\A_1$}
\put(50,6){$\A_{3}$}
\put(62,6){$\hat{\A}_{3}$}
\put(46,18){$\hat{\A}_2$}
\put(46,9){$\A_2$}
\put(31,2.5){\framebox(6,5){$\FF_{\A_1}$}}
\put(43,2.5){\framebox(6,5){$f$}}
\put(55,2.5){\framebox(6,5){$\FF_{\A_{3}}$}}
\put(43,11.5){\framebox(6,5){$\FF_{\A_2}$}}
\end{picture}
\caption{Fourier transform of multivariate function $f(a_1, a_2, a_3)$.}
\label{Fig19}
\end{figure}

More generally, given a complex-valued multivariate function $f(\ab)$ defined on a set of variables $\Ab = \{A_i\}$ whose alphabets $\A_i$ are vector spaces over $\F$, its Fourier transform is defined as the complex-valued function
$$
F(\hat{\ab}) = \sum_{\ab} f(\ab) \prod_i \omega^{\inner{\hat{a}_i}{a_i}}.
$$
In other words, in a normal factor graph, each variable $A_i$ may be transformed separately, as illustrated in Figure~\ref{Fig19}.  In \cite{AM10}, this property is called \emph{separability}.

Now let us define $U = V = \FF_{\A}$ and $S = \Phi_{\sim}/|\A|$, where the \emph{sign inverter indicator function} over $\hat{\A}$ is defined as
$$
\Phi_{\sim}(\hat{a}, \hat{a}')= \left\{
\begin{array}{ll} 
1,  & \mathrm{~if~} \hat{a} = -\hat{a}'; \\
0, & \mathrm{~otherwise}. 
\end{array} \right.
$$
Then the concatenation $U\!SV$ is the identity, since
$$
\sum_{\hat{a} \in \hat{\A}, \hat{a}' \in \A} \omega^{\inner{\hat{a}}{a}}\Phi_{\sim}(\hat{a}, \hat{a}')  \omega^{\inner{\hat{a}'}{a'}}  = \sum_{\hat{a} \in \hat{\A}}  \omega^{\inner{\hat{a}}{a - a'}} = |\A|\delta_{aa'},
$$
by a basic orthogonality relation for Fourier transforms over finite groups (see, \eg \cite{F98}).  This result is illustrated in Figure~\ref{Fig20}, where we omit the scale factor of $|\A|$.

\begin{figure}[h]
\setlength{\unitlength}{4pt}
\centering
\begin{picture}(50,4)(3, 3)
\put(0,5){\line(1,0){15}}
\put(7,6){$\A$}
\put(17,4){=}
\put(20,5){\line(1,0){5}}
\put(22,6){$\A$}
\put(25,2.5){\framebox(5,5){$\FF_{\A}$}}
\put(30,5){\line(1,0){5}}
\put(32,6){$\hat{\A}$}
\put(35,2.5){\framebox(5,5){$\Phi_{\sim}$}}
\put(40,5){\line(1,0){5}}
\put(42,6){$\hat{\A}$}
\put(45,2.5){\framebox(5,5){$\FF_{\A}$}}
\put(50,5){\line(1,0){5}}
\put(52,6){$\A$}
\end{picture}
\caption{A concatenation of factors that is equivalent to an edge, up to scale.}
\label{Fig20}
\end{figure}

Now we can prove our desired result:

\vspace{1ex}
\noindent
\textbf{Normal factor graph duality theorem} \cite{AM10, F11}. 
Given an NFG with partition function $Z(\xb)$, comprising external variables $X_i$ associated with half-edges, internal variables $Y_j$ associated with ordinary edges (all alphabets being vector spaces over a finite field $\F$), and complex-valued factors $f_k$ associated with vertices, the \textbf{dual normal factor graph} is defined by replacing each alphabet $\X_i$ or $\Y_j$ by its dual alphabet $\hat{\X}_i$ or $\hat{\Y}_j$, each factor $f_k$ by its Fourier transform $\hat{f}_k$, and finally by placing a sign inverter indicator function $\Phi_{\sim}$ in the middle of every ordinary edge.  Then the partition function of the dual NFG is the Fourier transform $\hat{Z}(\hat{\xb})$ of $Z(\xb)$, up to scale.\footnote{As shown in \cite{AM10}, the scale factor is $|\Y|$.}  \qed \vspace{1ex}

\noindent
\emph{Proof}:  Let us first convert the given NFG with partition function $Z(\xb)$ to an NFG with partition function $\hat{Z}(\hat{\xb})$, up to scale, by appending a Fourier transform $\FF_{\X_i}$ from $\X_i$ to $\hat{\X}_i$ to every half-edge associated with every external variable $X_i$, as in Figure~\ref{Fig19}.  Then let us replace every ordinary edge associated with every internal variable $Y_j$ by a concatenation $\FF_\A \Phi_\sim \FF_\A$ like that shown in Figure~\ref{Fig20};  this preserves the partition function $\hat{Z}(\hat{\xb})$, up to scale.  Now each vertex associated with each factor $f_k$ is surrounded by Fourier transforms of all of the variables involved in $f_k$, so it and its surrounding transforms may be replaced by a single vertex representing the Fourier transform factor $\hat{f}_k$ without changing the partition function, up to scale.  \qed \vspace{1ex}

Notice that this remarkably general theorem applies to any normal factor graph, whether or not it has cycles.

Using the fact that the indicator functions of  a linear code $\CC$ over $\F$ and of its orthogonal code $\CC^\perp$ are a Fourier transform pair, up to scale, one obtains as an immediately corollary a duality theorem for normal factor graph representations of linear codes  \cite{AM10, F11}, which is equivalent to the original normal graph duality theorem of \cite{F01}.

\section{Further Developments}

We now sketch briefly how the ``tree-based reparameterization'' approach of
Wainwright \etal \cite{WJW03}, the ``loop calculus'' results of Chertkov and
Chernyak \cite{CC06, CC06b}, and the Lagrange duality results of Vontobel and
Loeliger \cite{V02, VL03} fit within this generalized framework.  The full
developments will appear in a subsequent version of this paper.

\subsection{Tree-based reparameterization}

Wainwright, Jaakkola, and Willsky \cite{WJW03} have shown how the sum-product algorithm applied to general graphs with cycles can be understood as a tree-based reparameterization algorithm, where each round of the message-passing algorithm reparameterizes marginal distributions over simple subtrees consisting of a pair of vertices connected by an edge.  More generally, they consider iterative algorithms that reparameterize distributions over arbitrary cycle-free subtrees of the graph, particularly spanning trees.

Let $\Xb$  be a set of $m$ variables $X_i$ taking values $x_i$ in finite alphabets $\X_i$, and let $E$ be a set of pairs $(X_i, X_j)$ indicating which pairs of variables are connected.  Suppose that the corresponding graph with vertices $X_i$ and edges $(X_i, X_j) \in E$ is a tree (\ie cycle-free).  Finally, suppose that a probability distribution $p(\xb)$ over these variables can be expressed as
$$
p(\xb) \propto \prod_{1 \le i \le m} \psi_i(x_i) \prod_{(X_i,X_j) \in E} \psi_{ij}(x_i, x_j),
$$
where the functions $\psi_i(x_i)$ and $\psi_{ij}(x_i, x_j)$ depend only on the singleton variables $X_i$ and pairs $(X_i, X_j)$, respectively.  (By the Hammersley-Clifford theorem, this can always be done when $p(\xb)$ is a positive Markov random field over the graph.)

\begin{figure}[h]
\setlength{\unitlength}{4pt}
\centering
\begin{picture}(30,15)(2, -4)
\put(0,5){\line(1,0){5}}
\put(1,6){$\X_i$}
\put(6.5,5){\circle*{3}}
\put(6.5,6.5){\line(0,1){4}}
\put(5,10.5){\line(1,0){3}}
\put(5.5,11){$X_i$}
\put(6.5,-0.5){\line(0,1){4}}
\put(4,-5.5){\framebox(5,5){$\psi_{i}$}}
\put(7,1){$\X_i$}
\put(8,5){\line(1,0){5}}
\put(9,6){$\X_i$}
\put(13,2.5){\framebox(5,5){$\psi_{ij}$}}
\put(18,5){\line(1,0){5}}
\put(19,6){$\X_j$}
\put(24.5,5){\circle*{3}}
\put(24.5,6.5){\line(0,1){4}}
\put(23,10.5){\line(1,0){3}}
\put(23.5,11){$X_j$}
\put(24.5,-0.5){\line(0,1){4}}
\put(22,-5.5){\framebox(5,5){$\psi_{j}$}}
\put(25,1){$\X_j$}
\put(26,5){\line(1,0){5}}
\put(27,6){$\X_j$}
\end{picture}
\caption{Fragment of NFG representing a probability distribution on a tree.}
\label{FigTPR}
\end{figure}
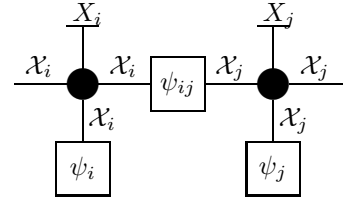

We can view such a distribution $p(\xb)$ as a partition function in which all variables are external (a ``global function'').  Normalizing this partition function, we obtain an equivalent partition function with the same external variables, but with an equality indicator function corresponding to each external variable replacing it in the corresponding normal factor graph.  A typical fragment of such an NFG is shown in Figure~\ref{FigTPR}.

Now we can execute the sum-product algorithm on such a cycle-free NFG, obtaining on each edge two messages, say $\muright{i}(x_{i})$ and $\muleft{i}(x_{i})$ on an edge with alphabet $\X_i$.  The corresponding marginal probability distribution $p_i(x_i)$ is proportional to the componentwise product of these messages:
$$p_i(x_i) \propto \muright{i}(x_{i}) \muleft{i}(x_{i}), x_i \in \X_i.$$

Such  a marginal distribution can be exhibited explicitly as a message in a ``reparameterized'' NFG by replacing a factor such as $\psi_{ij}(x_i, x_j)$ by the concatenation of three factors:
\begin{eqnarray*}
U(x_i, x_i') & = & \muleft{i}(x_{i}) \delta(x_i, x_i'); \\
S(x_i', x_j') & = & \frac{\psi_{ij}(x_i', x_j')}{\muleft{i}(x_{i}') \muright{j}(x_{j}')} \\
V(x_j, x_j') & = & \muright{j}(x_{j}) \delta(x_j, x_j'),
\end{eqnarray*}
which evidently preserves the partition function.

Such a reparameterization can be performed also in a graph with cycles, or over a subtree of a given graph.  Nice results are obtained when the messages are those that occur at a fixed point of the sum-product algorithm, but the messages do not have to be chosen in this way.  

In future work, we plan to use this approach to restate and generalize many of the results of \cite{WJW03} and related papers.

\subsection{Loop calculus}

Chertkov and Chernyak \cite{CC06, CC06b, CC09} have developed a ``loop calculus'' for statistical systems defined on finite graphs that allows the partition function of a system to be expressed as a finite sum over ``generalized loops,'' in which the lowest-order term corresponds to the Bethe-Peierls (sum-product algorithm) approximation.

We briefly sketch our approach to their results.
Suppose that all alphabets are binary.  Then replace every edge $Y_j$ in the system by the concatenation $U_jS_jV_j$, where in matrix notation
\begin{eqnarray*}
U_j & = & \left[\begin{array}{cc}
         +\muleft{j}(0)  & -\muright{j}(1) \\
         +\muleft{j}(1) & +\muright{j}(0) 
 \end{array} \right]; \\
S_j & = & \frac{1}{\Delta_j} \left[\begin{array}{cc}
         1  & 0 \\
         0 & 1
 \end{array} \right]; \\
V_j & = & \left[\begin{array}{cc}
         +\muright{j}(0)  & +\muright{j}(1) \\
         -\muleft{j}(1) & +\muleft{j}(0) 
 \end{array} \right], \\
\end{eqnarray*}
where $\muleft{j}(y_j)$ and $\muright{j}(y_j)$ are functions that may (but need not) be chosen as fixed-point messages of the sum-product algorithm, and $\Delta_j =
\muright{j}(0) \muleft{j}(0) + \muright{j}(1) \muleft{j}(1)$ is the determinant of
$U_j$ and $V_j$.  Evidently the concatenation $U_jS_jV_j$ is the identity, so this
replacement preserves the partition function.
 
Now express every $S_j$ as the sum of two matrices:
$$
S_j = \frac{1}{\Delta_j} \left[\begin{array}{cc}
        1  & 0 \\
        0 & 0
\end{array} \right] 
+
\frac{1}{\Delta_j} \left[\begin{array}{cc}
        0  & 0 \\
        0 & 1
\end{array} \right]; 
$$
if there are $n$ edges $Y_j$, then the partition function of the original NFG can
correspondingly be expressed as the sum of the partition functions of the
$2^n$ component NFGs.

If the functions $\muleft{j}(y_j)$ and $\muright{j}(y_j)$ are fixed-point
messages of the sum-product algorithm, then it turns out that the partition
function of the ``zero-order'' component graph is the Bethe-Peierls partition
function (at that fixed-point of the sum-product algorithm); that the
partition function of any component graph with a ``loose end'' (a vertex of
effective degree 1) is zero; and that the partition functions of the remaining
component graphs (corresponding to ``generalized loops,'' in which all
vertices have effective degree 2 or more) are ``small'' multiples of the
Bethe-Peierls partition function. Again, the full development will be given in a
subsequent version of this paper.
  
  \pagebreak
  
\subsection{Lagrange duality}
  
Structurally similar operations can be used to obtain the Lagrange duality
results for normal graphs of Vontobel and Loeliger \cite{V02, VL03}, which are
based on the Legendre transform of convex optimization theory.
  
One interesting aspect of this development is that instead of sums of
products, we consider minima over sums (\ie the sum-product semiring over the
reals $\R$ is replaced by the min-sum semiring over the extended real line
$\bar{\R} = \R \cup \{+ \infty\}$).  Thus a partition function has the following form:
$$
Z(\xb) = \min_{\yb \in \Y} \sum_{k \in \K} f_k(\xb_k, \yb_k), \quad \xb \in \X,
$$ 
where the ``factors'' $f_k(\xb_k, \yb_k)$ are $\bar{\R}$-valued.

The dual functions under the Legendre transform are functions in the max-sum
semiring.  Dualization involves the insertion of sign inverters into edges, as
with Fourier dualization.  Again, details will be provided in future versions
of this paper.
  
\section*{Acknowledgment}

We wish to acknowledge our close collaboration with Yongyi Mao, which
led to the normal factor graph paradigm.

\end{document}